\documentclass[twocolumn]{aastex701}

\usepackage{xspace}
\usepackage{booktabs}
\usepackage{multirow}
\usepackage{layouts}
\usepackage{amsmath}

\newcommand{\Vshell}[2]{\mathbf{#1}^\mathrm{#2}}
\newcommand{\shell}[2]{{#1}^\mathrm{#2}}

\newcommand{\va}[1][]{v_{\mathrm{A}#1}}

\newcommand{\V}[1]{\mathbf{#1}}

\newcommand{\cs}{c_{\mathrm{s}}}
\renewcommand{\Re}{\mathrm{Re}}
\newcommand{\Rm}{\mathrm{Rm}}
\newcommand{\Pm}{\mathrm{Pm}}

\newcommand{\T}[1]{\mathcal{T}_\mathrm{#1}}

\newcommand{\athenapk}{\texttt{AthenaPK}\xspace}

\usepackage[normalem]{ulem}

\begin{document}
\title{As a matter of reconstruction  -- 
varying the magnetic Prandtl number in implicit large eddy simulations
}
\shorttitle{Effective Pm in ILES}
\shortauthors{Kumar \& Grete}
\correspondingauthor{Vinay Kumar}
\email{vinay.kumar@icts.res.in}

\author[0000-0002-5610-8331]{Vinay Kumar}
\affiliation{%
International Centre for Theoretical Sciences, Tata Institute of Fundamental Research, Bangalore 560089, India
}
\email{vinay.kumar@icts.res.in}

\author[0000-0003-3555-9886]{Philipp Grete}
\affiliation{
Hamburger Sternwarte,
Universität Hamburg, Gojenbergsweg 112, 21029, Hamburg, Germany}
\email{pgrete@hs.uni-hamburg.de}

\begin{abstract}
The magnetic Prandtl number ($\Pm$), the ratio of viscosity to magnetic diffusivity,
spans many orders of magnitude in astrophysical plasmas. It shapes how turbulence
amplifies and structures magnetic fields. 
In direct numerical simulations (DNS) the viscosity and magnetic diffusivity can be controlled explicitly and independently, whereas in implicit large eddy simulations (ILES) they are set by the numerical scheme. We test whether, in a finite-volume
code, $\Pm$ can be varied through the reconstruction scheme by applying different methods to the hydro and magnetic variables. As a proof of concept, we simulate driven, subsonic, homogeneous magnetohydrodynamic turbulence using five pairs of hydrodynamic and magnetic reconstruction schemes at varying resolutions. We measure the effective
viscosity and resistivity from the energy transfer between scales.
Reconstructing the magnetic field at higher order than the fluid raises the
effective magnetic Prandtl number and vice versa.
We demonstrate
that a mix of second- and third-order reconstruction alone results in a $\gtrsim20\times$ change in $\Pm$.
A comparison of DNS with matched explicit dissipation at these values agrees
with the ILES. 
Mixed reconstruction thus offers a simple and efficient way to explore
$\Pm \not\simeq 1$ in ILES without adding explicit dissipation.
\end{abstract}


\section{Introduction}

Turbulence mediates the transport of energy, momentum, heat, and magnetic flux
in astrophysical plasmas, and it governs the evolution of many systems from the
interstellar medium to galaxy clusters \citep{elmegreen2004,brunetti2014}. In magnetized plasmas it is also widely
thought to generate and sustain cosmic magnetic fields through the turbulent
dynamo \citep[see][for reviews]{brandenburg2005, rincon2019}. A natural
starting point for understanding such flows is homogeneous, isotropic
turbulence: a statistically stationary, driven flow with no mean shear, no
boundaries, and no preferred direction, in which the physics of the cascade and
of dissipation can be studied in isolation.

Homogeneous isotropic turbulence in an \emph{unmagnetized} fluid is a one-parameter 
problem in the following sense. For a given forcing and at low Mach number, the
properties of the flow are controlled by the Reynolds number $\Re = UL/\nu$,
where $U$ and $L$ are a characteristic velocity and length scale and $\nu$ is
the kinematic viscosity. At large $\Re$ the larger scales (or the inertial range) become independent
of $\nu$, which only determines where the cascade terminates.
Homogeneous isotropic MHD turbulence is richer. The magnetic field is
dissipated separately, controlled via the magnetic diffusivity $\eta$, which
introduces a second control parameter, the magnetic Reynolds number
$\Rm = UL/\eta$. Equivalently, the problem can be described by $\Re$ and the
magnetic Prandtl number $\Pm = \Rm/\Re = \nu/\eta$, which sets the relative
position of the viscous and resistive dissipation scales. If no mean field is
imposed, the field strength is not an additional input but is set by the
dynamo itself. For a given forcing, homogeneous isotropic MHD turbulence is
therefore characterized by two numbers rather than one.

In astrophysical plasmas the magnetic Prandtl number is poorly constrained
and spans many orders of magnitude. It depends on the ionization fraction, temperature, and density of the gas,
ranging from $\Pm \sim 10^{-5}$--$10^{-2}$ in stellar convection zones to
$\Pm \gtrsim 10^{14}$ in the intracluster medium
\citep{brandenburg2005, schekochihin2004, brunetti2014}.
The consequences of non-unity $\Pm$ for turbulence and dynamo action have therefore
been explored extensively \citep[e.g.][]{brandenburg2005, rincon2019}. Much of
this work uses homogeneous isotropic MHD turbulence, which offers a clean
setting in which to isolate the effect of $\Pm$. With non-helical driving and
no mean field, the magnetic field is amplified and sustained by the
small-scale dynamo below the forcing scale. Both the growth rate of this
dynamo and the structure of the field it produces depend on $\Pm$
\citep{schekochihin2004, haugen2004, schekochihin2007, iskakov2007, bhat2013,
brandenburg23} because $\Pm$ sets the relative size of the viscous and
resistive scales on which the field is stretched and dissipated.

Astrophysical values of $\Re$ and $\Rm$ are enormous. Simulations therefore
aim for values as high as possible, so that a well-developed inertial range
forms.
The numerical resolution, however, sets the absolute limit of the dynamic range of a simulation. 
The maximum values of $\Re$ and $\Rm$ in practice are
even lower given finite resolution of the dissipative processes. 

In direct numerical
simulations (DNS), explicit viscosity and resistivity are included and the
dissipation scales are resolved, so $\Pm$ is chosen directly. In implicit
large eddy simulations (ILES), by contrast, no explicit dissipation is added,
and the truncation error of the numerical scheme removes energy close to the grid
scale. ILES can therefore be cheaper at a given resolution, because their time
step is limited only by the advective stability condition, without the
additional diffusive constraint of explicitly integrated dissipation terms.
They also reach higher effective $\Re$ and $\Rm$, because no grid points are
spent on resolving a larger dissipation range \citep{grete2023, shivakumar2025}. 
One drawback is that the dissipation is no longer under direct control: the effective viscosity and resistivity, 
and hence the effective $\Pm$, are inherited from the numerical scheme rather than chosen.

Finite-volume codes, however, offer a possible handle on this. Typically, they
reconstruct the hydrodynamic and magnetic variables using identical methods.
This reconstruction determines how strongly each field is smoothed near the grid
scale. However, the two fields can, in principle, receive different numerical
dissipation, and hence a different effective $\Pm$, through the choice of
reconstruction alone. In this paper, we test this idea by measuring the
effective viscosity, resistivity, and $\Pm$ of homogeneous, isotropic MHD
turbulence for different combinations of hydrodynamic and magnetic
reconstruction.

As a proof of concept, we perform simulations with \athenapk using five pairs of hydrodynamic and magnetic
reconstruction schemes, from donor-cell (first-order) to third-order WENO-Z, at four
resolutions between $128^3$ and $1024^3$. Raising the order of the magnetic
reconstruction relative to the hydrodynamic one raises $\Pm_\mathrm{eff}$, and
the reverse lowers it. At fixed forcing, Riemann solver, and time integrator,
the reconstruction pair alone changes $\Pm_\mathrm{eff}$ by a factor of about
21 at $N=1024$, from $0.26$ to $5.4$ across the four pairs for which it can be
measured reliably. This demonstrates that the reconstruction can be used to vary the magnetic Prandtl number of an ILES, reaching both the
sub-unity and super-unity regimes within a single code. We stress that this is
a proof of concept for a single code and a single flow configuration rather
than a general calibration. Section~\ref{sec:numerics} describes the
simulations and the energy transfer analysis used to estimate the dissipation
coefficients, Section~\ref{sec:results} presents the results, and
Section~\ref{sec:conclusions} discusses their implications and limitations.

\section{Numerical details}
\label{sec:numerics}

\subsection{Simulations}
\label{sec:sims}

\begin{deluxetable*}{ccc@{}c@{\hspace{2.5em}}cccccccc}
\tablecaption{Parameters of Our Simulations\label{tab:runs}}
\tablewidth{0pt}
\tablehead{
\multicolumn{3}{c}{Input Parameters} & \colhead{} & \multicolumn{8}{c}{Simulation Properties} \\
\cline{1-3} \cline{5-12}
\colhead{(1)} & \colhead{(2)} & \colhead{(3)} & \colhead{} &
\colhead{(4)} & \colhead{(5)} & \colhead{(6)} & \colhead{(7)} &
\colhead{(8)} & \colhead{(9)} & \colhead{(10)} & \colhead{(11)} \\
\colhead{Hydro} & \colhead{Mag.} & \colhead{$N$} & \colhead{} &
\colhead{$\mathcal{M}_{\rm s}$} & \colhead{$\mathcal{M}_{\rm A}$} &
\colhead{$\tau_{\rm e}$} & \colhead{$\eta_{\rm eff}$ [$10^{-6}$]} &
\colhead{$\nu_{\rm eff}$ [$10^{-6}$]} & \colhead{$\mathrm{Re}_{\rm eff}$} &
\colhead{$\mathrm{Rm}_{\rm eff}$} & \colhead{$\mathrm{Pm}_{\rm eff}$}
}
\startdata
\multirow{4}{*}{3P} & \multirow{4}{*}{2L} & 128 &  & 0.21 & 3.42 & 2.37 & $195\pm9$ & $48.3\pm3.4$ & 2180 & 539 & $0.247\pm0.012$ \\
 &  & 256 &  & 0.19 & 2.14 & 2.63 & $66.0\pm3.2$ & $17.2\pm1.1$ & 5510 & 1440 & $0.261\pm0.004$ \\
 &  & 512 &  & 0.18 & 1.80 & 2.73 & $25.0\pm0.5$ & $6.67\pm0.24$ & $13\,700$ & 3670 & $0.267\pm0.005$ \\
 &  & 1024 &  & 0.18 & 1.48 & 2.81 & $9.81\pm0.25$ & $2.57\pm0.07$ & $34\,600$ & 9070 & $0.262\pm0.006$ \\ \cline{1-12}
\multirow{4}{*}{2L} & \multirow{4}{*}{2L} & 128 &  & 0.19 & 2.35 & 2.58 & $153\pm9$ & $99.6\pm3.4$ & 972 & 632 & $0.650\pm0.030$ \\
 &  & 256 &  & 0.18 & 1.86 & 2.79 & $55.4\pm1.6$ & $41.1\pm1.3$ & 2180 & 1620 & $0.742\pm0.023$ \\
 &  & 512 &  & 0.17 & 1.65 & 2.86 & $20.8\pm0.2$ & $17.8\pm0.3$ & 4910 & 4210 & $0.857\pm0.021$ \\
 &  & 1024 &  & 0.17 & 1.30 & 2.99 & $8.04\pm0.04$ & $7.58\pm0.06$ & $11\,000$ & $10\,400$ & $0.943\pm0.012$ \\ \cline{1-12}
\multirow{4}{*}{2L} & \multirow{4}{*}{3P} & 128 &  & 0.17 & 1.61 & 2.91 & $47.9\pm2.9$ & $119\pm2$ & 725 & 1800 & $2.48\pm0.13$ \\
 &  & 256 &  & 0.16 & 1.34 & 3.15 & $15.0\pm0.5$ & $53.7\pm0.9$ & 1480 & 5290 & $3.58\pm0.07$ \\
 &  & 512 &  & 0.15 & 1.17 & 3.24 & $5.00\pm0.18$ & $23.7\pm0.4$ & 3250 & $15\,400$ & $4.73\pm0.08$ \\
 &  & 1024 &  & 0.16 & 1.14 & 3.15 & $1.90\pm0.05$ & $10.3\pm0.2$ & 7680 & $41\,800$ & $5.45\pm0.08$ \\ \cline{1-12}
\multirow{4}{*}{2L} & \multirow{4}{*}{3Z} & 128 &  & 0.17 & 1.56 & 2.88 & $55.5\pm3.1$ & $115\pm3$ & 753 & 1560 & $2.08\pm0.14$ \\
 &  & 256 &  & 0.16 & 1.44 & 3.06 & $19.0\pm0.8$ & $50.4\pm1.2$ & 1620 & 4300 & $2.65\pm0.05$ \\
 &  & 512 &  & 0.16 & 1.30 & 3.09 & $6.94\pm0.18$ & $21.8\pm0.5$ & 3700 & $11\,600$ & $3.15\pm0.06$ \\
 &  & 1024 &  & 0.16 & 1.20 & 3.10 & $2.73\pm0.03$ & $9.41\pm0.19$ & 8560 & $29\,500$ & $3.44\pm0.02$ \\ \cline{1-12}
\multirow{4}{*}{1C} & \multirow{4}{*}{3P} & 128 &  & 0.12 & 0.96 & 4.19 & $10.6\pm0.7$ & $513\pm13$ & 116 & 5600 & $48.2\pm3.9$ \\
 &  & 256 &  & 0.13 & 1.03 & 3.88 & $1.12\pm0.24$ & $277\pm8$ & 233 & $57\,600$ & $248\pm60$ \\
 &  & 512 &  & 0.13 & 0.96 & 3.87 & \nodata & $144\pm2$ & 448 & \nodata & \nodata \\
 &  & 1024 &  & 0.13 & 0.87 & 3.85 & \nodata & $76.2\pm2.3$ & 851 & \nodata & \nodata \\
\enddata
\tablecomments{The hydrodynamic and magnetic reconstruction schemes and the resolution $N$ are inputs; all other columns show time-averaged quantities measured in the saturated state. Sonic and
Alfv\'enic rms Mach numbers $\mathcal{M}_{\rm s}=u_{\rm rms}/c_{\rm s}$ and
$\mathcal{M}_{\rm A}=u_{\rm rms}/v_{\rm A,rms}$; eddy turnover time
$\tau_{\rm e}=L_{\rm f}/u_{\rm rms}$ with $L_{\rm f}=2\pi/k_{\rm f}=L/2$ (in units of
$L/c_{\rm s}$); $\mathrm{Re}_{\rm eff}=u_{\rm rms}L_{\rm f}/\nu_{\rm eff}$,
$\mathrm{Rm}_{\rm eff}=u_{\rm rms}L_{\rm f}/\eta_{\rm eff}$, and
$\mathrm{Pm}_{\rm eff}=\nu_{\rm eff}/\eta_{\rm eff}$. $\mathrm{Re}_{\rm eff}$ and $\mathrm{Rm}_{\rm eff}$ are rounded to three
significant figures. The dissipation coefficients are in units of $c_{\rm s}L$ and fitted for
$kL/2\pi\geq8$; the uncertainties are statistical only. For \texttt{1C-3P} at $N\ge512$ the fitted resistivity is not
positive, so $\eta_{\rm eff}$, $\mathrm{Rm}_{\rm eff}$ and $\mathrm{Pm}_{\rm eff}$ are not given.}
\end{deluxetable*}

We solve the equations of ideal compressible MHD using the \athenapk code\footnote{\athenapk is available and maintained at
\url{https://github.com/parthenon-hpc-lab/athenapk}}, which is built on top of the performance-portable
adaptive-mesh refinement framework \texttt{Parthenon} \citep{parthenon} in a triply periodic cubic box of side $L=1$
discretized on a uniform grid of $N^3$ cells. Fluxes are computed with the
HLLD Riemann solver and time integration uses the second-order van Leer (VL2)
predictor--corrector scheme with a CFL number of $0.3$. Divergence errors in
the magnetic field are controlled using a mixed hyperbolic-parabolic Generalized
Lagrange Multiplier (GLM) cleaning scheme \citep{dedner2002}, which advects
and damps $\nabla \cdot \mathbf{B}$.

Reconstruction to the cell interfaces is performed on the primitive variables,
separately for the hydrodynamic set $(\rho,\, \mathbf{u},\, p)$ and for the
magnetic field $\mathbf{B}$, using the five scheme pairs listed in
Table~\ref{tab:runs}. The GLM scalar $\psi$ is reconstructed with the same
scheme as $\mathbf{B}$.\footnote{We verified for a subset of runs that the
reconstruction of $\psi$ has no appreciable effect on the results. This is
possibly because $\nabla\cdot\mathbf{B}$ errors remain small in
weakly compressible flows, such as our subsonic runs.}  We label each pair by the nominal order and family of the scheme applied to the
hydrodynamic and magnetic variables respectively: \texttt{1C} denotes
first-order piecewise-constant (donor cell, DC), \texttt{2L} second-order
piecewise-linear (PLM), \texttt{3P} third-order piecewise-parabolic (PPM), and
\texttt{3Z} third-order WENO-Z. Thus \texttt{2L-3P} denotes PLM reconstruction
of the hydrodynamic variables combined with PPM reconstruction of the magnetic
field. The set spans a fiducial, matched-order pair (\texttt{2L-2L}), as well as pairs in which the two
sets of variables are reconstructed at different orders, in both directions. We define $\Delta(\text{order})$ as the difference between the nominal order
of the magnetic reconstruction and that of the hydrodynamic reconstruction, so
that $\Delta(\text{order})>0$ corresponds to the magnetic field being
reconstructed at higher order than the hydrodynamic variables, and
$\Delta(\text{order})<0$ to the reverse.
Each pair is run at four resolutions, $N = 128,\ 256,\ 512$ and $1024$, giving
$20$ simulations in total, which allows the dependence on the scheme pair to
be separated from the dependence on grid resolution.

We adopt an adiabatic index $\gamma = 1.0001$ to maintain approximate
isothermality. The initial state is a uniform plasma at rest,
$\mathbf{u} = 0$, with density $\rho = 1$ and pressure $p = 1$,
giving an approximately isothermal sound speed
$\cs = \sqrt{\gamma p/\rho} \approx 1$. 
The combination of $L=1$, $\rho=1$, and $\cs=1$ determines the unit system
of the simulations. Lengths are in units of $L$, densities in units of the
initial density, and velocities in units of $\cs$. Times are then in units of
$L/\cs$, and the diffusivities $\nu$ and $\eta$ are in units of $\cs L$.

The plasma is threaded by a weak,
zero-net-flux magnetic field
\begin{equation}
    B_x = B_{\rm amp}\sin(2\pi z/L), \qquad B_y = B_z = 0,
    \label{eq:binit}
\end{equation}
expressed in Heaviside--Lorentz units, so that the magnetic pressure is
$B^2/2$, the plasma beta is $\beta = 2p/B^2$, and the Alfv\'{e}n speed is
$\va = B/\sqrt{\rho}$. The amplitude $B_{\rm amp} = \sqrt{2p/\beta_0}$ is
set by an initial plasma beta $\beta_0 = 2\times 10^{4}$. 

Turbulence is driven by a stochastic, purely solenoidal acceleration field which evolves as an Ornstein--Uhlenbeck process \citep{eswaran1998,schmidt2009, grete2018} with correlation time $t_{\rm corr} = 1.0\, L/\cs$. 
The acceleration field has an inverse parabolic profile in wavenumber space, supported over $1 \le kL/2\pi \le 3$ and peaked at $k_{\rm f} = 2\times 2\pi/L$, and its amplitude is normalized to give a fixed root-mean-square acceleration $a_{\rm rms} = 0.1\, \cs^2/L$ in each simulation. The resulting flow is subsonic in all runs, with the sonic Mach number
$\mathcal{M}_{\rm s} = u_{\rm rms}/\cs \leq 0.25$ at all times. 

The seed field is dynamically unimportant throughout the initial transient,
and is subsequently amplified by turbulent dynamo action. Each run is therefore a small-scale
dynamo problem, a configuration in which $\Pm$ is known to influence
the evolution of the magnetic field
\citep[e.g.][]{schekochihin2004, haugen2004, schekochihin2007}. 
Our analysis is done over 15 snapshots in a window that starts well after the dynamo has saturated, when the kinetic and magnetic energies are approximately stationary.
This window spans $3.3$--$5.9$
large-eddy turnover times ($\tau_{\rm e} = 2\pi/k_{\rm f} u_{\rm rms} $) depending on the run. All runs start from identical initial conditions and use the same forcing
parameters and random seed, so that systematic differences between them can be
attributed primarily to the reconstruction scheme.

\subsection{Energy transfer analysis and effective dissipation coefficients}
\label{sec:transfer}

Since the simulations contain no explicit dissipation, the viscosity and
resistivity they realize must be measured \textit{a posteriori}. We do this with the
shell-to-shell energy transfer framework of \citet{grete2017}, extended to
include explicit dissipative terms by \citet{grete2023}, and summarize it here briefly.

The velocity and magnetic fields are decomposed into logarithmically spaced shells in Fourier space. A superscript denotes the shell to
which a field has been filtered, so that $\Vshell{u}{K}$ is the velocity
restricted to the wavenumbers in shell $K$, and likewise $\Vshell{B}{Q}$ for
the magnetic field. Throughout, we use the mass-weighted velocity
$\V{w} = \sqrt{\rho}\,\V{u}$, so that the spectral kinetic energy density
based on $\tfrac{1}{2}w^2$ is positive definite \citep{kida1990}. The transfer
functions
\begin{equation}
    \T{XY}(Q,K), \qquad
    \mathrm{X},\mathrm{Y} \in \{\mathrm{U},\mathrm{B}\} ,
    \label{eq:transfer}
\end{equation}
then measure the rate at which energy is moved from shell $Q$ of budget
$\mathrm{X}$ into shell $K$ of budget $\mathrm{Y}$, with $\mathrm{U}$ and
$\mathrm{B}$ denoting the kinetic and magnetic energy budgets respectively. The subscript thus records
both the source and the destination budget: $\T{UU}$ is the kinetic cascade,
$\T{BB}$ the magnetic cascade, and $\T{UB}$ a transfer from the kinetic to the
magnetic budget. Where a transfer between budgets proceeds through more than
one mechanism the mediating process is appended, so that $\T{UBT}$ and
$\T{UBP}$ denote kinetic-to-magnetic transfer via magnetic tension and
magnetic pressure respectively. Two further terms carry no second budget
index: $\T{PU}$, the internal-to-kinetic transfer via the pressure gradient,
and $\T{FU}$, the injection of energy by the mechanical forcing. Collecting
the channels that feed a given shell, the scale-wise kinetic and magnetic
energy budgets read
\begin{align}
    \partial_t \shell{E}{K}_{\rm kin} &= \int \sum_Q
        \bigl( \T{UU} + \T{BUT} + \T{BUP}
        \nonumber \\
        &\hphantom{{}=\int \sum_Q \bigl(}
        + \T{PU} + \T{FU} \bigr) \mathrm{d}\V{x} + \mathcal{D}_{\rm U}^K ,
        \label{eq:balkin} \\[6pt]
    \partial_t \shell{E}{K}_{\rm mag} &= \int \sum_Q
        \left( \T{BB} + \T{UBT} + \T{UBP} \right)
        \mathrm{d}\V{x} + \mathcal{D}_{\rm B}^K , \label{eq:balmag}
\end{align}
where $\mathcal{D}_{\rm U}^K$ and $\mathcal{D}_{\rm B}^K$ collect the 
numerical dissipation acting on shell $K$ of each budget.

In a simulation with explicit viscosity and resistivity, the dissipative
terms take the same shell-filtered form as the other transfers,
\begin{align}
    \T{\nu U}(Q,K) &= \nu \int \sqrt{\rho}\,\Vshell{w}{K} \cdot
        \left( \Delta \Vshell{u}{Q} + \tfrac{1}{3}\nabla
        \left( \nabla \cdot \Vshell{u}{Q} \right) \right) \mathrm{d}\V{x} ,
        \label{eq:tnu} \\
    \T{\eta B}(Q,K) &= \eta \int \Vshell{B}{K} \cdot
        \Delta \Vshell{B}{Q}\, \mathrm{d}\V{x} ,
        \label{eq:teta}
\end{align}
and enter Equations~(\ref{eq:balkin}) and (\ref{eq:balmag}) alongside the
other transfers. All terms that change the energy of a shell are then
included explicitly. The budgets therefore close without a remainder, and
the residual terms $\mathcal{D}_{\rm U}^K$ and $\mathcal{D}_{\rm B}^K$ vanish. In an ILES, Equations~(\ref{eq:tnu}) and (\ref{eq:teta}) have no counterpart
in the equations solved: dissipation is supplied by the truncation error of
the scheme and is not available as a transfer function. 

For stationary
turbulence the left-hand sides of Equations~(\ref{eq:balkin}) and
(\ref{eq:balmag}) vanish on average, since the system is in global balance, so
the sum of the explicitly represented transfers into a given shell must be
balanced by the dissipation acting on it,
\begin{equation}
    \mathcal{D}_{\rm U}^K = - \int \sum_Q
        \left( \T{UU} + \T{BUT} + \T{BUP} + \T{PU} + \T{FU} \right)
        \mathrm{d}\V{x} ,
    \label{eq:resid}
\end{equation}
and correspondingly for $\mathcal{D}_{\rm B}^K$. Every transfer in the sum is
measured directly from the simulation data, so the residual is obtained
without any assumption about the form of the numerical dissipation. It is this
measured, scale-wise residual to which we fit Equations~(\ref{eq:tnu}) and
(\ref{eq:teta}), with $\nu$ and $\eta$ as the free parameters.

The residuals are evaluated for each snapshot in the analysis window and
time-averaged. Since $\T{\nu U}$ and $\T{\eta B}$ are linear in $\nu$ and
$\eta$, each fit reduces to a one-parameter linear least-squares problem.
We perform it over wavenumbers $kL/2\pi \ge k_{\min} = 8$ to exclude the
forcing band.\footnote{The sensitivity of the fitted values to $k_{\min}$ is
discussed in Appendix~\ref{app:kmin}.}
Uncertainties are estimated by splitting the analysis window into three
equal sub-windows and taking the largest departure of the sub-window values
from the full-window fit, and are propagated into $\Pm = \nu/\eta$.

\section{Results}
\label{sec:results}

In all simulations, the volume-averaged kinetic and magnetic energies evolve through the same sequence of stages. Initially, the forcing spins up the flow, and the weak seed field is amplified rapidly by the small-scale dynamo while remaining dynamically unimportant (kinematic stage). Once the magnetic energy is large enough, the Lorentz force begins to back-react on the flow, which ends the kinematic stage and reduces the kinetic energy from its peak value. The magnetic energy continues to grow during this nonlinear stage, but at a much slower, non-exponential rate. Finally, the dynamo saturates, and both energies fluctuate around quasi-stationary values. Our analysis window begins well after saturation.

\subsection{Field structure and energy spectra}
\label{sec:spectra}

\begin{figure*}
\centering
\includegraphics[width=\textwidth]{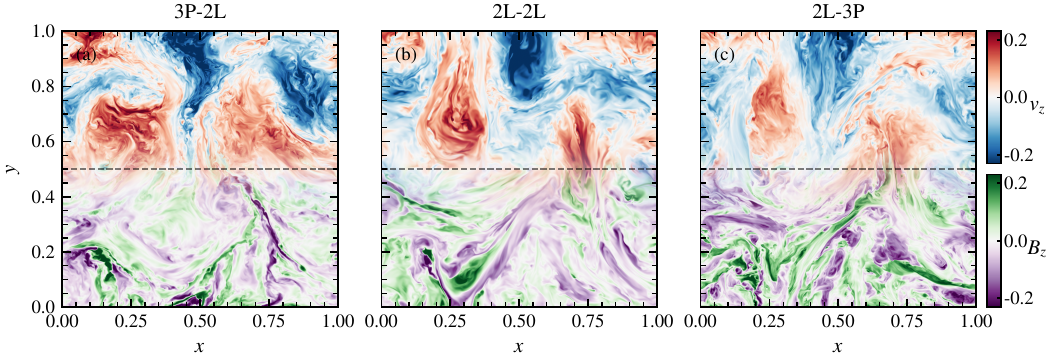}
\includegraphics[width=\textwidth]{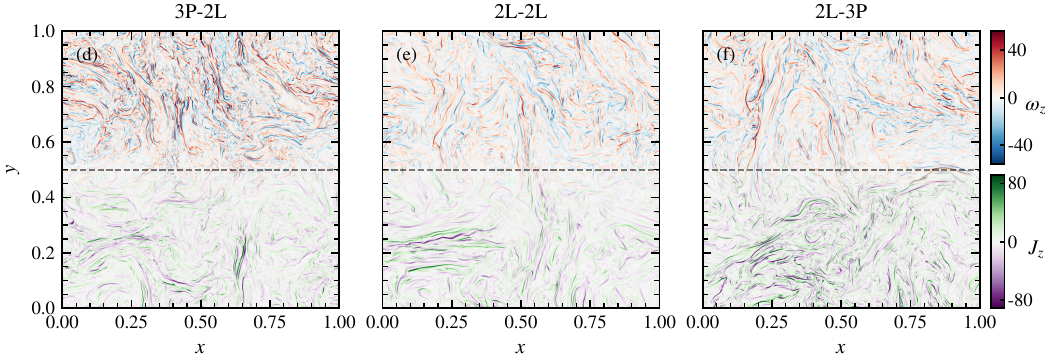}
\caption{Constant-$z$ slices at a representative time for the $N=1024$ runs \texttt{3P-2L},
\texttt{2L-2L}, and \texttt{2L-3P} (left to right). Panels (a)--(c) show the
out-of-plane components of the velocity, $u_z$, and the magnetic field, $B_z$.
Panels (d)--(f) show the out-of-plane components of the vorticity, $\omega_z$
(upper half), and the current density, $J_z$ (lower half). In each panel the
dashed line separates the two fields. The color scales of $u_z$ and $B_z$ are
common to all panels and clipped at the 60th percentile of $|u_z|$ and
$|B_z|$, respectively. The color scales of $\omega_z$ and $J_z$ are also
common to all panels but are not clipped, so they span the full range of the
data.}
\label{fig:fields}
\end{figure*}

Figure~\ref{fig:fields} shows slices of $u_z$ and $B_z$ (panels a--c) and of
the vorticity and current density, $\omega_z$ and $J_z$ (panels d--f), at
$N=1024$ for \texttt{3P-2L}, \texttt{2L-2L}, and \texttt{2L-3P}, taken in the
saturated state. In all three runs, the velocity is dominated by structures at
large scales, while the magnetic field is organized into thin, elongated
filaments and folds. The differences between the reconstruction pairs are
modest in $u_z$ and $B_z$. There is a hint that the velocity field of
\texttt{3P-2L} contains more fine-scale structure than in the other two runs,
and, likewise, that the magnetic field of \texttt{2L-3P} has the most filamentary content.

The differences are easier to see in $\omega_z$ and $J_z$. Being derivatives
of the velocity and magnetic fields, they emphasize small scales. The vorticity of \texttt{3P-2L} is visibly more intense and more
fine-scale than in \texttt{2L-2L} and \texttt{2L-3P}, which are smoother and
weaker. Conversely, the fine-scale structure in $J_z$ is most pronounced for
\texttt{2L-3P}, whereas \texttt{2L-2L} and \texttt{3P-2L} show smoother current sheets.
These trends agree with the impressions from
$u_z$ and $B_z$, but they are based on a single slice at a single time and
remain qualitative. They are, however, in the direction suggested by the
spectra discussed below.

\begin{figure*}
\centering
\includegraphics[width=\textwidth]{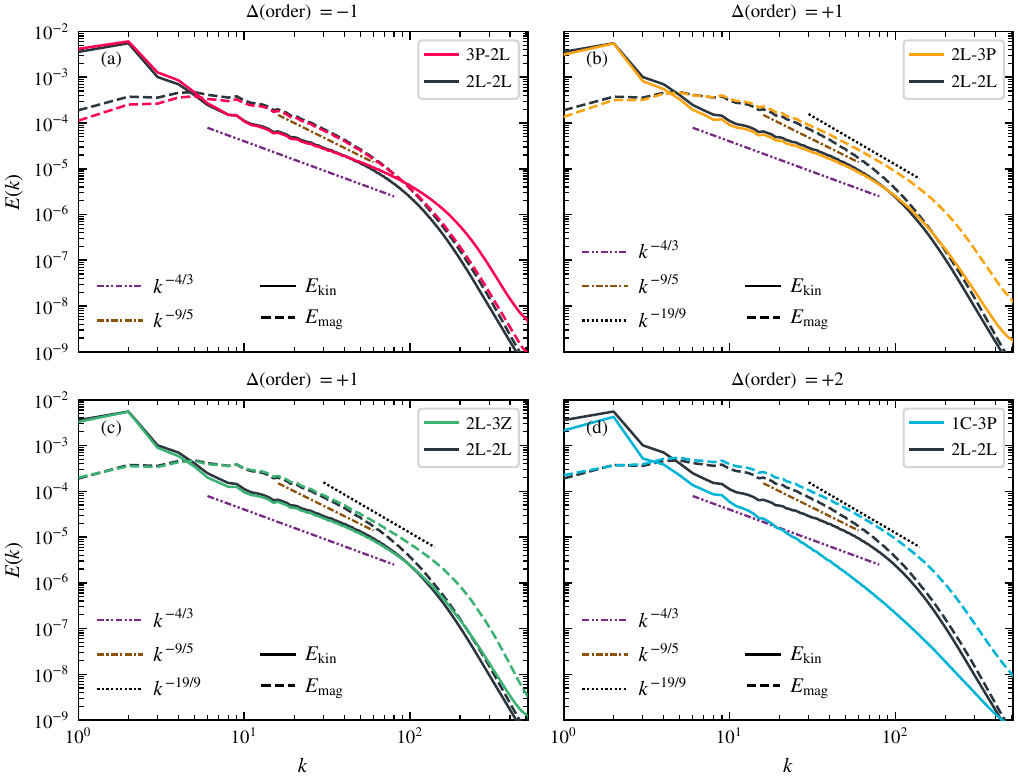}
\caption{Time-averaged kinetic (solid) and magnetic (dashed) energy spectra
at $N=1024$. Each panel compares one reconstruction pair with the
matched-order reference \texttt{2L-2L}; panels are ordered by
$\Delta(\text{order})$. The wavenumber axis is in units of $2\pi/L$ and has been truncated at the Nyquist wavenumber.}
\label{fig:spectra}
\end{figure*}

Figure~\ref{fig:spectra} shows the time averaged kinetic and magnetic energy spectra at $N=1024$. Each panel compares one
reconstruction pair with the reference \texttt{2L-2L}. For the reference
pair, the kinetic and magnetic spectra are damped at visually similar scales.
Changing the reconstruction of only one of the two fields predominantly
changes the damping scale of the corresponding spectrum. In panel (a), the
magnetic spectrum of \texttt{3P-2L} closely matches that of \texttt{2L-2L},
as both use the same magnetic reconstruction. Its kinetic spectrum extends
to larger $k$ because of the higher-order hydrodynamic reconstruction. The
opposite holds in panels (b) and (c). The kinetic spectra of \texttt{2L-3P}
and \texttt{2L-3Z} follow that of \texttt{2L-2L}, apart from a small
extension that may indicate a ``reverse dynamo'' \citep{brandenburg2019}. Their
magnetic spectra extend to larger $k$. 
Finally, in panel (d), the first-order
hydrodynamic reconstruction of \texttt{1C-3P} strongly damps the kinetic
spectrum, which falls off at much smaller $k$ than in all other runs. Its
magnetic spectrum extends at least as far as in the other
$\Delta(\text{order})>0$ pairs.

We also show reference scalings in Figure~\ref{fig:spectra}. For the kinetic
spectra, we show $k^{-4/3}$, a slope found in earlier ILES of subsonic MHD
turbulence and attributed to the suppression of the Kolmogorov-like cascade
($\T{UU}$) by magnetic tension \citep{grete2021,grete2023}. For the magnetic spectra, we
show $k^{-9/5}$, which was found empirically by \citet{fielding2023} and
\citet{beattie2025} in simulations that used the same second-order scheme for
all variables, and which therefore correspond most closely to our matched-order
pair \texttt{2L-2L}. For $\Delta(\text{order})>0$, we also show $k^{-19/9}$,
the magnetic spectrum that \citet{galishnikova2022} predicted for the 
$\Pm\geq1$ dynamo, in which the field forms folds that become unstable to tearing at high
$\Rm$.
Below the tearing scale, the spectrum follows from the theory of
tearing-mediated turbulence \citep{mallet2017,loureiro2017}. Our dynamic range is too
limited to establish a power law, so these lines serve as visual guides only,
although the $\Delta(\text{order})>0$ runs hint at a steepening similar to
$k^{-19/9}$. 

Taken together, the spectra indicate that the reconstruction order of each
field sets the wavenumber at which that field is dissipated. Where the
magnetic spectrum extends to larger $k$ than the kinetic one, the magnetic
field is dissipated at smaller scales than the flow, which suggests
$\Pm_\mathrm{eff}>1$. The reverse suggests $\Pm_\mathrm{eff}<1$. This is the
case for $\Delta(\text{order})>0$ and $\Delta(\text{order})<0$, respectively.
For the matched-order
pair \texttt{2L-2L}, the two spectra are damped at similar scales, and we
expect $\Pm_\mathrm{eff}$ of order unity. We quantify $\Pm_\mathrm{eff}$
below.

\subsection{Effective dissipation coefficients and $\Pm_\mathrm{eff}$}
\label{sec:coeffs}

\begin{figure*}
\centering
\includegraphics[width=\textwidth]{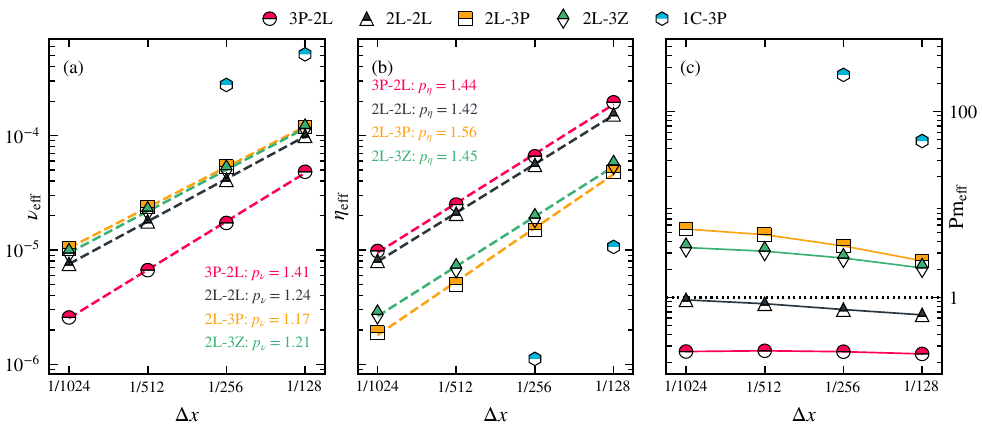}
\caption{Effective kinematic viscosity $\nu_\mathrm{eff}$ (left), effective
resistivity $\eta_\mathrm{eff}$ (middle), and effective magnetic Prandtl
number $\Pm_\mathrm{eff}=\nu_\mathrm{eff}/\eta_\mathrm{eff}$ (right) as
functions of grid spacing $\Delta x$ for all reconstruction pairs. Lines in the left
and middle panels are power-law fits, with the exponents, $p_\nu$ and $p_\eta$, given
in the panels. For \texttt{1C-3P}, the points at $N=512$ and $1024$ are
excluded (see text). The horizontal dotted line in the right panel denotes $\Pm_{\rm eff} = 1$.}
\label{fig:nueta}
\end{figure*}

Figure~\ref{fig:nueta} shows the effective coefficients obtained from the
energy transfer analysis (Section~\ref{sec:transfer}). For all pairs, both
$\nu_\mathrm{eff}$ and $\eta_\mathrm{eff}$ decrease with increasing
resolution, as expected for ILES, where the dissipation is set by the
truncation error of the scheme. The decrease follows empirical power laws,
$\nu_\mathrm{eff}\propto\Delta x^{p_\nu}$ and
$\eta_\mathrm{eff}\propto\Delta x^{p_\eta}$, with the fitted exponents listed
in the respective panels. The unreliable \texttt{1C-3P} points are excluded
from the fits (see below). At fixed $N$, changing the reconstruction of one
field mainly changes its own diffusivity, with the higher-order scheme giving
the lower value.

The resulting $\Pm_\mathrm{eff}$ is listed in Table~\ref{tab:runs} and shown in
the right panel of Figure~\ref{fig:nueta}. It is ordered by
$\Delta(\text{order})$. At $N=1024$ we measure $\Pm_\mathrm{eff}=0.26$ 
for \texttt{3P-2L}, $0.94$ for \texttt{2L-2L}, $3.4$ for
\texttt{2L-3Z}, and $5.4$ for \texttt{2L-3P}. The reconstruction
pair alone thus changes $\Pm_\mathrm{eff}$ by a factor of $\sim21$, at fixed
forcing, Riemann solver, and time integrator. This confirms the trend
suggested by the spectra. The pairs with $\Delta(\text{order})>0$ have
$\Pm_\mathrm{eff}>1$ and \texttt{3P-2L} has $\Pm_\mathrm{eff}<1$. 
The two pairs with $\Delta(\text{order})=+1$ differ by about $60\%$, hinting at an additional
effect from the reconstruction method itself\footnote{WENO-Z is formally fifth-order.}.
The matched-order pair \texttt{2L-2L} stays just below unity and increases towards it with increasing resolution.

For \texttt{1C-3P}, the fits are poor. The effective resistivity is not
positive definite at $N\geq512$, and we therefore exclude these points. At
$N=128$ and $256$, the fitted resistivity depends strongly on the fit range,
so the corresponding values are unreliable and we do not interpret them. We
list them in Table~\ref{tab:runs} for completeness only. The spectra in panel (d)
of Figure~\ref{fig:spectra} nevertheless suggest $\Pm_\mathrm{eff}>1$ for this
pair, although we cannot quantify this with the present method.

\subsection{Comparison with DNS}
\label{sec:dns}

The energy transfer analysis assigns effective coefficients to an ILES. It
does not show that a simulation with these coefficients as explicit
dissipation would be equivalent to the ILES. We test this below.

We take \texttt{2L-3P} at $N=256$. Its effective coefficients are
$\nu_\mathrm{eff}=5.37\times10^{-5}$ and $\eta_\mathrm{eff}=1.50\times10^{-5}$
in units of $\cs L$ (Table~\ref{tab:runs}). They correspond to
$\Re_\mathrm{eff}\approx1.5\times10^{3}$, $\Rm_\mathrm{eff}\approx5.3\times10^{3}$,
and $\Pm_\mathrm{eff}=3.58$. We rerun this setup at $N=512$ and $1024$ with
these values imposed as constant explicit viscosity and resistivity. The DNS use the same initial conditions, forcing, random seed, Riemann
solver, and time integrator as the ILES. They use third-order parabolic reconstruction in all variables (\texttt{3P-3P} in our notation).

\begin{figure*}
\centering
\includegraphics[width=\textwidth]{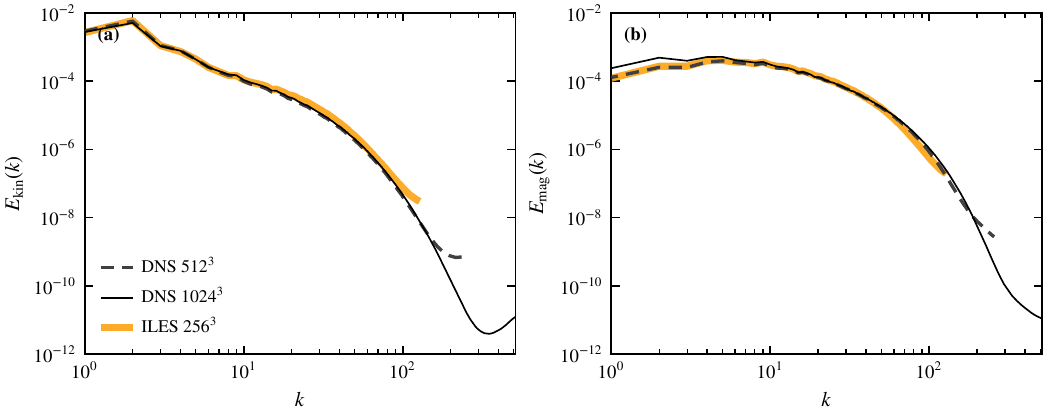}
\caption{Time-averaged kinetic (left) and magnetic (right) energy spectra of
the \texttt{2L-3P} ILES at $N=256$ and of DNS at $N=512$ and $1024$. The DNS
use $\nu=\nu_\mathrm{eff}$ and $\eta=\eta_\mathrm{eff}$ measured from the
$N=256$ ILES. Each spectrum ends at the Nyquist wavenumber of its run. The axis
ends at that of the $N=1024$ DNS.}
\label{fig:dns}
\end{figure*}

Figure~\ref{fig:dns} compares the time-averaged spectra of the $256^3$ ILES with those of the DNS.
Over the wavenumber range of the ILES, the kinetic and magnetic spectra of
both DNS follow those of the ILES closely. We repeated the test for the
$128^3$ \texttt{2L-3P} run, with DNS at $N=512$ and $1024$ using its coefficients
(Table~\ref{tab:runs}); the spectra agree in the same way. 
Explicit dissipation with $\nu_{\rm eff}$ and $\eta_{\rm eff}$ thus reproduces the
spectra of both ILES, so a scalar viscosity and resistivity describe well their
numerical dissipation at this level. The effective coefficients are also consistent with an independent
residual-based estimate (Appendix~\ref{app:pymhd}).

\section{Discussion and Conclusions}
\label{sec:conclusions}

We asked whether the choice of reconstruction scheme can vary the effective
magnetic Prandtl number of an implicit large eddy simulation. 
We simulated
driven, subsonic, homogeneous MHD turbulence with \athenapk, using five pairs
of hydrodynamic and magnetic reconstruction at resolutions from $128^3$ to
$1024^3$. 
We measured the effective viscosity and resistivity from the energy
transfer between scales.

The answer is yes, within the scope of these runs. Reconstructing the magnetic
field at higher order than the fluid raises $\Pm_\mathrm{eff}$, and the reverse
lowers it. DNS with the
measured coefficients reproduce the ILES spectra for the runs we tested.

This may matter for small-scale dynamo studies in cosmological simulations.
Many of these rely on numerical dissipation
\citep[e.g.][]{martinalvarez2022, pakmor2017}, so their effective magnetic
Prandtl number is set by the scheme rather than chosen and is usually not controlled.
Recent code comparisons find that the amplification of the magnetic field
differs between numerical schemes, and attribute this to differences in
numerical diffusivity \citep{pakmor2026, ugalino2026}. In our homogeneous,
isotropic flow, the choice of reconstruction changes the effective magnetic
Prandtl number without any explicit dissipation. Whether the same holds in a
cosmological setting, with its shocks and self-gravity, we have not tested.

The magnetic Prandtl number also matters beyond dynamos. In plasmoid-mediated
reconnection, the reconnection rate has been found to depend on it
\citep{comisso2015}, but this has recently been challenged \citep{kumar2026}.
The same question arises in the decay of magnetically dominated turbulence,
where energy is transferred inversely from small to large scales even without
net magnetic helicity \citep{zrake2014, brandenburg2015}. This inverse transfer has
recently been analyzed with shell-to-shell transfer functions
\citep{kasselmann2026}. If the decay is limited by reconnection rather than by
the Alfv\'en time, its timescale could carry an additional dependence on Pm
\citep{hosking2023}. Simulations find that the decay time exceeds the Alfv\'en
time by a factor that grows with the Lundquist number
\citep{bhat2021, brandenburg2024}. In two dimensions this factor levels off at
large Lundquist numbers, but this cannot yet be confirmed in three dimensions
\citep{brandenburg2024}. Whether a Pm dependence survives at the large Pm and
small diffusivities of the early Universe is also still debated
\citep{brandenburg2024, anandavijayan2026}. An ILES used for such problems would
inherit its Pm from the scheme in the same way, so the choice of reconstruction
could matter there too.

Accretion disks are another case where Pm matters. In local simulations of the
magnetorotational instability, the turbulent stress increases with Pm
\citep{fromang2007}, and simulations without explicit dissipation have an
effective Pm set by the scheme \citep{simon2009}. The same holds for global
disks, where numerical diffusivity can limit dynamo growth \citep{nixon2024},
even though global simulations find large-scale dynamo action
\citep[e.g.][]{santhiya2026}.

Our results also suggest a possible route to a wider range in
$\Pm_\mathrm{eff}$. Among the pairs we tested, $\Pm_\mathrm{eff}$ increases with
$\Delta(\text{order})$, spanning a factor of $\sim21$ at $N=1024$
(Table~\ref{tab:runs}). If this trend continues, reconstruction schemes of
higher order than those used here, in particular in the magnetic field, could
extend this range. We have not tested this. It is also unclear whether
$\Pm_\mathrm{eff}$ depends on the order difference alone or also on the absolute
orders of the two reconstructions. Testing schemes beyond those considered here
would settle this.

Our study is a proof of concept, with one code, one flow, and one forcing. 
The value of $\Pm_\mathrm{eff}$ is set by the scheme pair. It drifts with resolution,
but by at most a factor of 2.2 between $N=128$ and $1024$, compared with a
factor of about 21 between pairs. The measured values can therefore guide the
choice of pair and resolution for a desired combination of
$\mathrm{Re}_\mathrm{eff}$ (or $\mathrm{Rm}_\mathrm{eff}$) and $\Pm_\mathrm{eff}$.
The effective Reynolds numbers remain far below astrophysical values, but within
this regime the reconstruction scheme is a practical handle on
$\Pm_\mathrm{eff}$ in ILES.

\begin{acknowledgments}
V.K. thanks P. S. Santhiya and C. Rajarshi for useful discussions.
V.K. acknowledges support from the Department of Atomic Energy, Government of India, under Project Nos. RTI4013 and RTI4019.
P.G. acknowledges funding by the Deutsche Forschungsgemeinschaft (DFG) -- 555983577.
This project received access to JUPITER through the JUPITER Research and Early Access Program (JUREAP). JUPITER is funded by the EuroHPC Joint Undertaking, the German Federal Ministry of Research, Technology and Space, and the Ministry of Culture and Science of the German state of North Rhine-Westphalia.
\texttt{AthenaPK} and \texttt{Parthenon} use the \texttt{Kokkos} library \citep{9485033}. Analysis and plotting routines made use of the packages \texttt{yt} \citep{2011ApJS..192....9T}, \texttt{NumPy} \citep{harris2020array}, \texttt{scipy} \citep{2020SciPy-NMeth} and \texttt{Matplotlib} \citep{Hunter:2007}.

\section*{Use of AI tools}
We used Claude (Anthropic) as a writing assistant while preparing this
manuscript. It was used to edit and tighten the text. The simulations, the analysis, the results and their
interpretation are the work of the authors. The authors checked all text and all
references, and take full responsibility for the content of this paper.
\end{acknowledgments}

\appendix

\section{Sensitivity of the fitted coefficients to fitting range}
\label{app:kmin}

\begin{figure*}
\centering
\includegraphics[width=\textwidth]{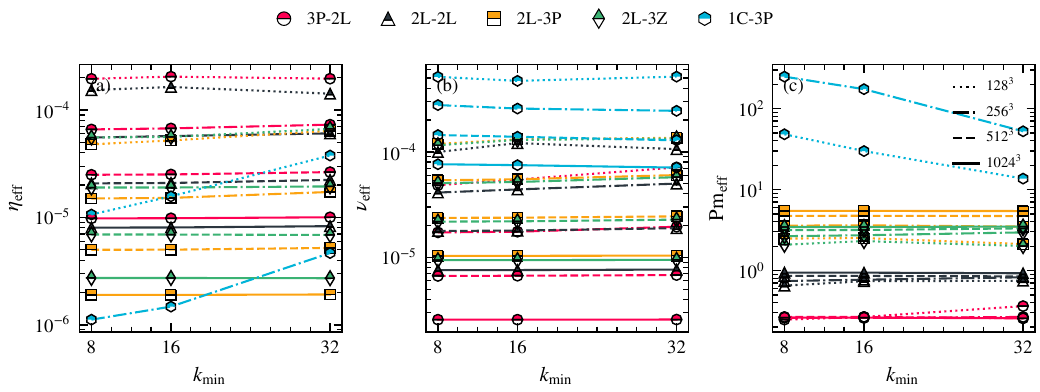}
\caption{Dependence of the fitted effective resistivity $\eta_\mathrm{eff}$
(a), effective viscosity $\nu_\mathrm{eff}$ (b), and effective magnetic
Prandtl number $\Pm_\mathrm{eff}$ (c) on the lower wavenumber bound
$k_{\min}$ of the fit range. Colors and markers denote the reconstruction
pair and line styles the resolution (legend in panel c). For \texttt{1C-3P},
$\eta_\mathrm{eff}$ and $\Pm_\mathrm{eff}$ are shown only for $N=128$ and
$256$ (see Section~\ref{sec:coeffs}).}
\label{fig:kmin}
\end{figure*}

The coefficients in Table~\ref{tab:runs} are fitted for $kL/2\pi \ge k_{\min}=8$,
which excludes the forcing band but is otherwise an arbitrary choice.
Figure~\ref{fig:kmin} repeats the fits for $k_{\min}=8$, 16, and 32; the values
at $k_{\min}=8$ are those of Table~\ref{tab:runs}.

Although the figure is crowded, most curves are nearly horizontal, so
$\nu_\mathrm{eff}$, $\eta_\mathrm{eff}$, and hence $\Pm_\mathrm{eff}$ depend
weakly or not at all on $k_{\min}$ for \texttt{3P-2L}, \texttt{2L-2L},
\texttt{2L-3P}, and \texttt{2L-3Z}. The exception is \texttt{1C-3P}, for which
$\eta_\mathrm{eff}$ changes by a factor of a few over this range, and
$\Pm_\mathrm{eff}$ by a comparable factor in the opposite direction, while
$\nu_\mathrm{eff}$ is much less affected. This agrees with our finding that
the resistivity of this pair is not reliably determined
(Section~\ref{sec:coeffs}). Since the uncertainties in Table~\ref{tab:runs}
do not include the $k_{\min}$ dependence, the conclusions for the other four
pairs rest on its being weak, as seen here.

\section{Cross-check of the effective coefficients with PyMHD}
\label{app:pymhd}

We cross-check the spectral-fit coefficients (Table~\ref{tab:runs}) for the
\texttt{2L-2L} and \texttt{2L-3P} runs. For this, we use the residual method of
\citet{hua2026}, implemented in the \texttt{PyMHD}\footnote{\url{https://github.com/PlasmaHua/PyMHD}} package.

In this method, the numerical dissipation is the residual of the momentum
and induction equations. We evaluate it on the simulation output with a
higher-order reconstruction (TENO7-M, $C_T=10^{-2}$). The time derivative is
also of high order and uses consecutive snapshots. We apply the method to
$N=128$, $256$ and $512$, with 10 consecutive snapshots in the saturated
state.

We convert the numerical dissipation rates,
$\mathbf{u}\cdot\mathbf{D}^{\rm num}_{\rm vis}$ and
$\mathbf{B}\cdot\mathbf{D}^{\rm num}_{\rm res}$, to energy-equivalent
coefficients,
\begin{equation}
  \nu_{\rm eff}=\frac{\langle\mathbf{u}\cdot\mathbf{D}^{\rm num}_{\rm vis}\rangle}
                     {\langle\mathbf{u}\cdot\nabla\!\cdot\!\mathbb{T}\rangle},
  \qquad
  \eta_{\rm eff}=\frac{\langle\mathbf{B}\cdot\mathbf{D}^{\rm num}_{\rm res}\rangle}
                      {\langle\mathbf{B}\cdot\nabla^{2}\mathbf{B}\rangle}.
  \label{eq:energy_equiv}
\end{equation}
Here, $\mathbb{T}$ is the viscous stress tensor for unit viscosity. The
brackets denote the average over the domain and the analyzed snapshots. An
explicit viscous (resistive) term with these coefficients removes the same
mean energy per unit time as the scheme. The coefficients are therefore
global energy measures. They do not imply that the numerical dissipation is
locally of Laplacian form.

\begin{figure*}
\centering
\includegraphics[width=\textwidth]{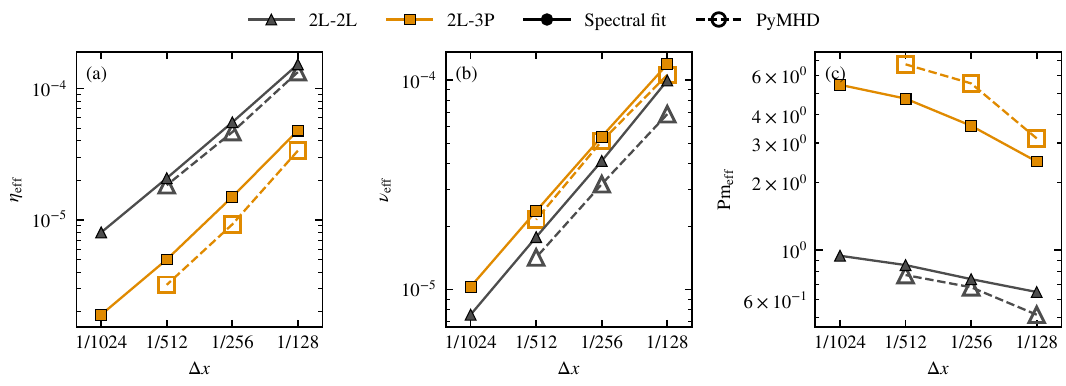}
\caption{Effective resistivity (a), viscosity (b) and magnetic Prandtl
number (c) versus grid spacing for \texttt{2L-2L} and  \texttt{2L-3P}. Solid:
spectral fits (Table~\ref{tab:runs}). Dashed: energy-equivalent coefficients
from the residual estimate with PyMHD \citep{hua2026}
(Equation~\ref{eq:energy_equiv}), available for $N\le512$.}
\label{fig:pymhd}
\end{figure*}

Figure~\ref{fig:pymhd} compares the two estimates. Both estimates show the same scaling with resolution between $N=128$ and
512. We find $\eta_{\rm eff}\propto\Delta x^{1.4}$ for \texttt{2L-2L} and
$\Delta x^{1.6\text{--}1.7}$ for \texttt{2L-3P}. For $\nu_{\rm eff}$, we find
$\Delta x^{1.1\text{--}1.2}$ in both cases. The ordering of the schemes is
also the same. At $N=512$, $\mathrm{Pm}_{\rm eff}$ of \texttt{2L-3P} exceeds that of
\texttt{2L-2L} by a factor of 5.5 (spectral fit) and 8.7 (residual).

Differences of this size are not unexpected. The two estimators weight scales
differently, and numerical dissipation is not strictly of Laplacian form
\citep{hua2026}. Despite these differences, both estimators show the same increase of $\mathrm{Pm}_{\rm eff}$ with a higher-order
magnetic reconstruction.

\bibliographystyle{aasjournal}
\bibliography{refs}

\end{document}